\documentclass[3p,times,procedia]{elsarticle}
\usepackage{nupha_ecrc}
\usepackage{bm, amsmath, color}

\volume{00}

\firstpage{1}

\journalname{Nuclear Physics A}

\runauth{}

\jid{nupha}

\jnltitlelogo{Nuclear Physics A}

\usepackage{amssymb}

\usepackage[figuresright]{rotating}

\newcommand\p{\partial}

\begin{document}

\begin{frontmatter}



\dochead{XXVIth International Conference on Ultrarelativistic Nucleus-Nucleus Collisions\\ (Quark Matter 2017)}

 \title{
Chiral magnetohydrodynamics for heavy-ion collisions
 }


\author{Yuji Hirono}

\address{
Department of Physics, Brookhaven National Laboratory,
Upton, New York 11973-5000
}
 

\begin{abstract}
The chiral magnetic effect (CME) is a macroscopic transport effect
resulting from the chiral anomaly. 
We review the recent progress in theoretical understanding the 
properties of chiral plasmas, in which the CME and other anomaly-induced 
 transports take place.
In particular, the nontrivial interplay of anomalous currents and
 dynamical electromagnetic fields is discussed. 
We also review the theoretical status of the modeling of anomalous
transport effects in heavy-ion collisions. 
\end{abstract}


\end{frontmatter}


\section{Introduction}
\label{intro}

Macroscopic transport effects arising from the chiral anomaly have been
attracting much attention in recent years.
For example, 
magnetic fields generate dissipationless electric currents when they are applied
to chirally imbalanced media, in which the 
numbers of left-handed and right-handed fermions are different. 
This is the chiral magnetic effect (CME)
\cite{Kharzeev:2007jp,Fukushima:2010vw}. 
Theoretically, the existence of CME can be derived by a number of ways
such as the perturbation 
theory, lattice QCD \& QED simulations \cite{Yamamoto:2011gk,
Bali:2014vja}, and holography \cite{Yee:2009vw}.
Moreover, 
the CME and other chiral transport effects constitute an integral part
of relativistic hydrodynamics.
Those effects are not only allowed but required from the consistency
with the second law of thermodynamics \cite{Son:2009tf}. 
Recently, the first experimental observation of CME using  a Dirac semimetal 
is reported \cite{Li:2014bha}. 
Heavy-ion collisions offer the opportunity to observe anomalous
transport effects as well.

In this talk, we would like to discuss the interplay of anomalous chiral effects and
{\it dynamical} electromagnetic fields.
The CME currents are generated by applied magnetic fields. 
The currents in turn produce 
electromagnetic fields, that affect the configuration of the
electromagnetic fields.
We are interested in the fate of such coupled systems of chiral media
and electromagnetic fields. 
For the complete understanding of various phenomena in such systems, 
one needs a consistent framework to describe both the chiral plasma and
the electromagnetic fields. Chiral magnetohydrodynamics (MHD) is such a
theory.
An ordinary MHD is a low-energy theory for electrically conducting fluids. 
It can describe the time evolution of the coupled system of the
conducting fluids and electromagnetic fields in a consistent way. 
In chiral MHD, the fluid is a chiral one, which includes the anomalous
chiral effects like CME as a medium response. 
It has been pointed out that the chiral plasma 
develops an instability \cite{Joyce:1997uy,Akamatsu:2013pjd}. 
Chiral MHD can answer the eventual fate of the instability. 
Description in terms of chiral MHD is appropriate and important not only for the heavy-ion
collisions, but also for  early Universe before electroweak phase
transition.
For example, the interplay of chiral fermions with dynamical gauge 
fields leads to a formation of large-scale magnetic fields like we see
in the current Universe. 

In this contribution, we review 
the recent theoretical progress understanding the nature of chiral
plasmas. 
We also review recent hydrodynamic attempts at describing CME in
heavy-ion collisions.

\section{Anomalous chiral effects and dynamical electromagnetic fields}

\subsection{Chiral anomaly and the topology of magnetic fields}

Let us start by explaining a relation of the topology of magnetic fields and fermions. 
The conservation of the axial current $j^\mu_A$ is broken by a quantum
effect, the extent which is quantified by the anomaly equation,
\begin{equation}
 \p_\mu j^{\mu}_A = C_A \bm E \cdot \bm B,
  \label{eq:anomaly}
\end{equation}
where $\bm E$ and $\bm B$ are electric and magnetic fields.
After a spatial integration,
the anomaly equation (\ref{eq:anomaly}) takes the following form, 
\begin{equation}
 \frac{d}{dt} \left[
\mathcal H + \mathcal H_F
	     \right] = 0 ,
\label{eq:total-helicity}
\end{equation}
where we have defined 
\begin{equation}
 \mathcal H \equiv \int d^3x \bm A \cdot \bm B  , \quad \quad
 \mathcal H_F \equiv \frac{2}{C_A} \int d^3 x \ n_A . 
\end{equation}
Here $\bm A$ is the vector potential, and $n_A$ is the axial charge
density.
We have introduced two helicities: 
$\mathcal H$ is so-called magnetic helicity, and
$\mathcal H_F$ is the total fermionic helicity.
Equation (\ref{eq:total-helicity}) tells us that 
the total ``chirality'' is a constant, although 
it can be stored either in magnetic fields or in fermions. 
When the chirality is stored in the fields, the field takes a topologically
nontrivial form.
Indeed, it is well known that
the magnetic helicity is a measure of the
topology of magnetic fields.
When the system is made of magnetic flux tubes, the magnetic helicity
can be written in terms of topological invariants as 
\begin{equation}
 \mathcal H 
  = \sum_{i} \mathcal S_i \varphi^2_i
   + 2 \sum_{i,j} \mathcal L_{ij} \varphi_i \varphi_j , 
\end{equation}
where
$\varphi_i$ is the magnetic flux of the $i-$th flux tube, 
$\mathcal S_i$ is the C\u{a}lug\u{a}reanu-White self-linking number,
and $\mathcal L_{ij}$ is the Gauss linking number 
\cite{moffatt1990energy,ricca1992helicity,moffatt1992helicity}.
The integrated anomaly equation (\ref{eq:total-helicity}) tells us that
fermions can change the
magnetic helicity, hence the topology of $\bm B$ fields.

\subsection{Inverse cascade}

Let us find how the fermions affect the $\bm B$ field topology. 
Coupled system of chiral matter and electromagnetism have been
studied using the Maxwell-Chern-Simons theory
\cite{Boyarsky:2011uy,Tashiro:2012mf,Manuel:2015zpa,Hirono:2015rla,Xia:2016any}.
The total helicity is conserved and this constrains the dynamics of the
system. 
Remarkably, such systems exhibit the so-called ``inverse cascade,''
in which the energy is transferred from smaller to larger scales. 
This leads to large structures of magnetic fields as the system
evolves. 
It also turns out that the evolution is self-similar
\cite{Hirono:2015rla}, although the existence of such solution might
depend on the choice of equation of state \cite{Tuchin:2017vwb}. 

This discussion can be extended to include the degrees of freedom of the fluid.
Fluid can also share the helicity in the form of fluid
helicity, $\int d^3 x \ \bm v \cdot \bm \omega$, where $\bm \omega
\equiv \nabla \times \bm v$ is the vorticity of the fluid. 
The turbulent spectrum in chiral MHD is discussed
\cite{Yamamoto:2016xtu,Pavlovic:2016gac} and self-similar inverse
cascade remains. 

\begin{figure}[htbp]
\center{
\includegraphics[width=0.6\textwidth]{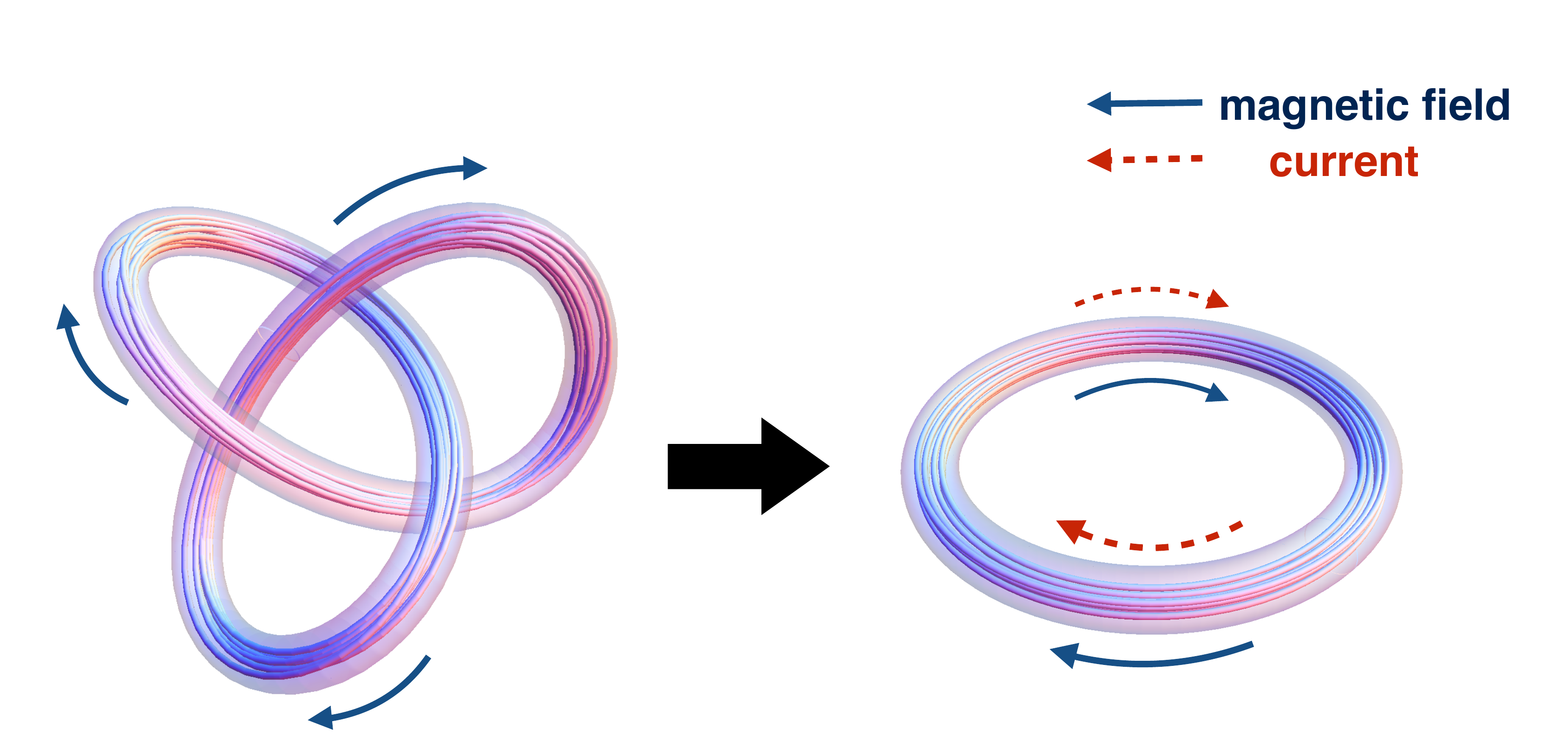}
 \caption{
 Illustration of the CME current generation from the change of topology
 of magnetic fields. 
}
}
 \label{fig:c-gen}
\end{figure}

\subsection{Quantized CME}

An explicit formula that connects the
change in the topology of $\bm B$ field and CME current
has been derived recently \cite{Hirono:2016jps}, 
\begin{equation}
\sum_{i} \oint_{C_i} \Delta \bm J \cdot d \bm x = - \frac{e^3}{2 \pi^2}
\Delta \mathcal H ,
\label{eq:deltaj}
\end{equation}
where $\Delta \bm J$ is the generated CME current, 
$C_i$ are the trajectories of the magnetic flux tubes, 
$\Delta \mathcal H$ is the change in magnetic helicity, 
and $e$ is the electric charge. 
Equation (\ref{eq:deltaj}) indicates that a change in the topology of
the magnetic fields ($\Delta \mathcal H$) necessarily results in
the generation of a CME current (see Fig. 1 as a sketch). This relation
can be extended to include vortices \cite{Kirilin:2017tdh}. 
Namely, the reconnections of magnetic fields and vortices can also lead
to the generation of CME currents in a chiral fluid.

\subsection{Equations of motion of (chiral) MHD}

Let us turn to the dynamical description of a chiral plasma.
A low-energy theory of a conducting fluid and electromagnetic fields
is MHD, equations of motions of which are given by
\footnote{
The formulation of MHD has recently been revisited in
Refs.~\cite{Grozdanov:2016tdf,Hernandez:2017mch}
}
\begin{equation}
 \p_\mu T^{\mu\nu}_{\rm tot}  = 0 ,
  \quad \p_\mu \tilde F^{\mu\nu} =0. 
\end{equation}
where $T^{\mu\nu}_{\rm tot}$ is the energy-momentum tensor of whole system, and
the latter equation is the Bianchi identity. 
Ideal MHD is characterized by the constitutive relation for the electric
field, $E^\mu_{(0)} = 0$. 
This corresponds to the limit of large conductivity.
$E^\mu$ is the electric field in the frame of the fluid element,
so the observer on the fluid element does not feel any electric field. 

In ideal MHD, the magnetic helicity is conserved and the topology of
$\bm B$ fields is unchanged, which can be seen as follows. 
The magnetic helicity is the volume integral of the zero-th component of
the Chern-Simons current $ h^\mu_B = \tilde F^{\mu\nu} A_\nu$. 
The divergence of $h^\mu_B$ reads 
\begin{equation}
\p_\mu  h^\mu_B =8 E_\mu B^\mu , 
\end{equation}
and since $E^\mu = 0$ in ideal MHD,  $h^\mu_B$ is conserved in this limit. 
Although anomalous chiral effects are dissipationless, they do not
appear in ideal MHD.

Anomalous effects enter if one considers the contribution from finite
conductivity $\sigma$. 
The correction from resistivity to the electric field reads 
\begin{equation}
 E^\mu_{(1)} =
  \lambda \epsilon^{\mu\nu\alpha\beta}
  \p_\alpha \left[ u_\nu B_\beta \right]
- \epsilon_B B^\mu 
- \epsilon_\omega \omega^\mu  ,
\label{eq:1st-order-e}
\end{equation}
where $\lambda \equiv 1/\sigma$ is the resistivity,
$\epsilon_B = \sigma_B / \sigma$, and 
$\epsilon_\omega = \sigma_\omega / \sigma$.
Coefficients $\sigma_B$ and $\sigma_\omega$ are chiral magnetic/vortical
conductivities. 
The latter two terms in Eq. (\ref{eq:1st-order-e}) are anomalous
effects.

\subsection{Linear excitations in chiral MHD}

In idea MHD, the magnetic field is ``frozen in'' to the fluid.
The fluid is pierced by magnetic fields and they move together. 
Because of the tension of the magnetic field and the moment of inertia
of the fluid, 
oscillatory motion of the magnetic field line happens, and it 
propagates along the magnetic field.
This is the Alfven wave.
The nature of the Alfven wave is affected by anomalous chiral
effects. 
If we take the wave vector parallel to $\bm B$,
the dispersion relation reads
\begin{equation}
 \omega =
  \pm v_A k_{||} 
  - \frac{i}2 \left[
	       \left(\bar \eta + \lambda \right)k_{||}^2
	       - s \epsilon_B k_{||}
	      \right], 
\end{equation}
where the Alfven velocity $v_A$ is defined by $v_A^2 \equiv B^2 /
(e+p+B^2)$,
$\bar \eta  \equiv \eta / (e+p+\bm B^2)$ is a normalized shear
viscosity, 
and $s$ is the helicity of the wave (there are left-handed
and right-handed Alfven waves). 
Because of the contribution proportional to $\sigma_B$,
helicity-dependent instability appears.
For example, if $\sigma_B>0$, the positive helicity modes with $k<k_c$
is unstable, where
\begin{equation}
 k_{\rm c} = \frac{\sigma_B}{1  + \bar \eta \sigma}. 
\end{equation}
This instability generates helical flows in the presence of chirality
imbalance, hence is a mechanism to transfer helicity from fermions to
fluid flow \cite{cmhd-h}.

\section{Anomalous hydrodynamic modeling of heavy-ion collisions}

\subsection{Chirality production and CME in the glasma }


Let us turn to the anomalous chiral effects in heavy-ion collisions. 
Shortly after the collisions, the matter is in a nonequilibrium state 
called glasma, which consists of highly occupied gluons.
Then, fermions are created from those fields and the system will reach 
the local equilibrium state described by hydrodynamics. 
In the context of the CME search, 
the glasma dynamics is important, because chromo $\bm E^a \cdot \bm B^a$
creates chirality imbalance, that is necessary for CME
current to be generated \cite{Tanji:2016dka}. 
The amount of axial charge in the initial stage of hydrodynamics 
strongly affect the value of the final observable, and its quantitative
estimation is important for the experimental detection of CME. 

The chirality generation from nonequilibrium color fields has been studied
via real-time classical lattice simulations. 
In Ref.~\cite{Mace:2016svc}, the sphaleron rate is measured in a
nonequilibrium non-Abelian plasma, and it is 
found to be enhanced
compared to the equilibrium values.
This means that a glasma can more efficiently produce chirality imbalance
than an equilibrium plasma. 

\begin{figure}[htbp]
\center{
\includegraphics[width=0.6\textwidth]{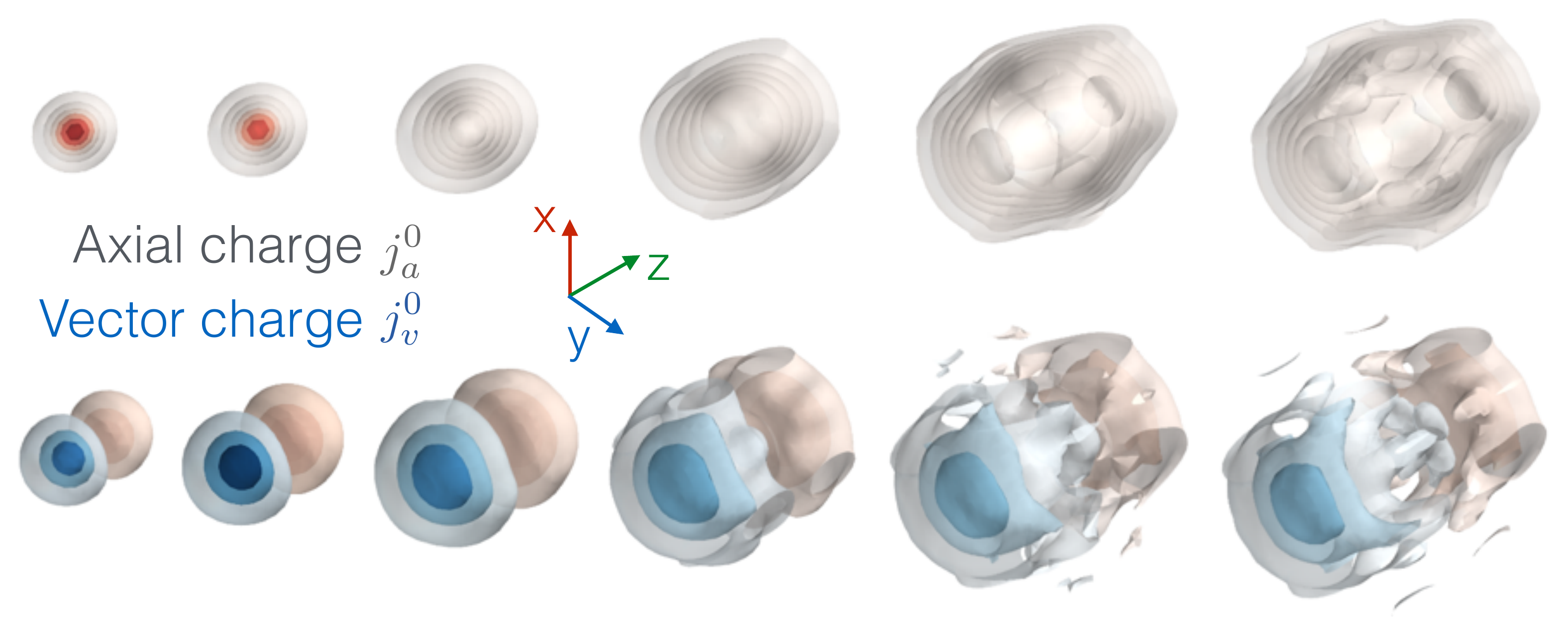}
 \caption{
Time evolution of the axial and vector charge densities. Taken from Ref.~\cite{Mueller:2016ven}. 
 }
}
 \label{fig:sphaleron}
\end{figure}

Furthermore, CME itself can also be happening in non-equilibrium states
of glasma and fermions. 
Indeed, CME is simulated in real-time lattice simulations,
in which $U(1)$ magnetic fields are applied in addition to
color fields \cite{Mueller:2016ven,Mace:2016shq}.
Figure 2
shows the time evolution of axial and vector
charge densities. 
A sphaleron transition creates axial
charges, and later CME current develops in the direction of the applied
magnetic field. 
Since magnetic fields are stronger at earlier times,
the contribution of the CME current in glasmas can be important for 
experimental search of CME.
Those pre-hydro CME currents should enter in the initial conditions of 
the subsequent anomalous hydrodynamic stage. 
In those works, the backreaction from the generated fermions to the
fields is not included, but it can also be incorporated.
In Ref.~\cite{Buividovich:2015jfa}, such a study is performed in the
case of an Abelian plasma.

\subsection{Anomalous hydrodynamic calculations}

To reach a decisive conclusion about the existence of anomalous chiral
effects in heavy-ion collisions \cite{sorensen2017},
we need a tool to describe this phenomena quantitatively.
For this purpose, hydrodynamic models with anomaly-induced transports
have been developed \cite{Hirono:2014oda,Yin:2015fca}.
%
\begin{figure}[htbp]
\center{
\includegraphics[width=0.35\textwidth]{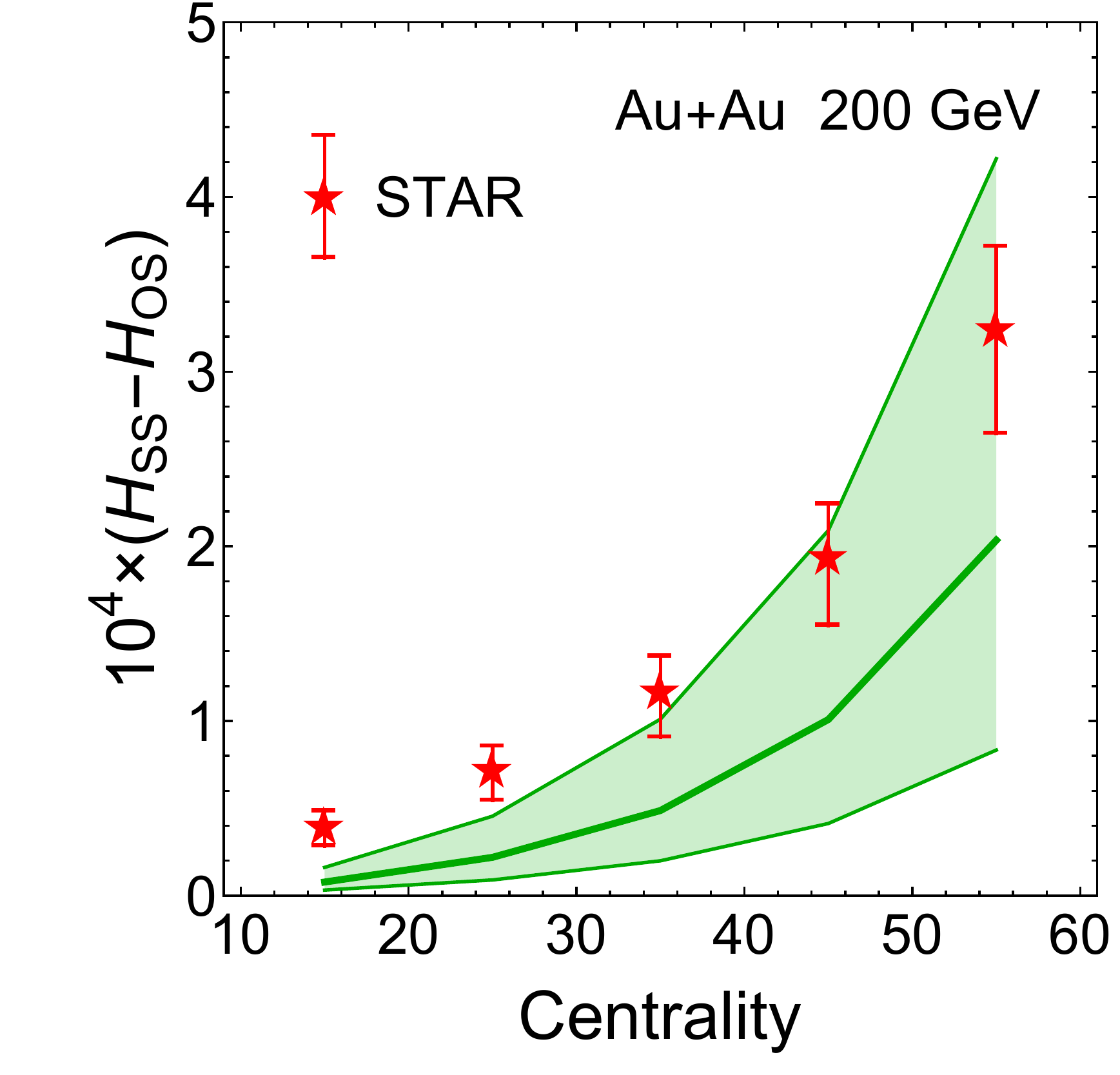}
\caption{Centrality dependence of H correlation. Taken from Ref.~\cite{Jiang:2016wve}. }
}
 \label{fig:cme-h}
\end{figure}
In Ref.~\cite{Jiang:2016wve},
charge transport from anomalous currents are studied on a background
solution of second order viscous fluid in 2+1 dimensions (VISHNew).
The authors incorporated the effects of resonance decays, which contribute as a
background effect, and simulations are performed on an event-by-event
basis.
Figure 3
shows \cite{Jiang:2016wve} the centrality
dependence of the so-called $H$ correlation \cite{bzdak2013charge},
which shows a very similar trend with the STAR data. 

Another approach for describing heavy-ion collisions is
chiral kinetic theory \cite{Son:2012wh,Stephanov:2012ki,Son:2012zy,wang2016}. 
Examples for such calculations have been reported\cite{Sun:2016mvh,Huang:2017tsq} in this Quark Matter.

\subsection{MHD and magnetic fields}

In the search of CME in heavy-ion collisions, 
one of the biggest uncertainties is the strength and life time of
magnetic fields.
The MHD-type description is also useful in investigating 
electromagnetic properties of the plasma
\cite{Hirono:2012rt,Gursoy:2014aka}. 
Recently, MHD simulations for heavy-ion collisions are performed
\cite{Inghirami:2016iru,Das:2017qfi} and the effects on observables
like $v_2$ is discussed. 
In Fig.~4,
the values of magnetic fields are
plotted as a function of Bjorken time \cite{Inghirami:2016iru}.
Compared to vacuum evolution (dotted line), the result from MHD
simulation (solid line) shows slower decay as an effect of the
medium.
The initial $\bm B$ field for the MHD calculation (which starts from
$\tau = 0.4 {\rm fm}$) is given by the
solution of Maxwell equations at finite conductivity \cite{Tuchin:2013apa}. 
\begin{figure}[htbp]
 \center{
\includegraphics[width=0.5\textwidth]{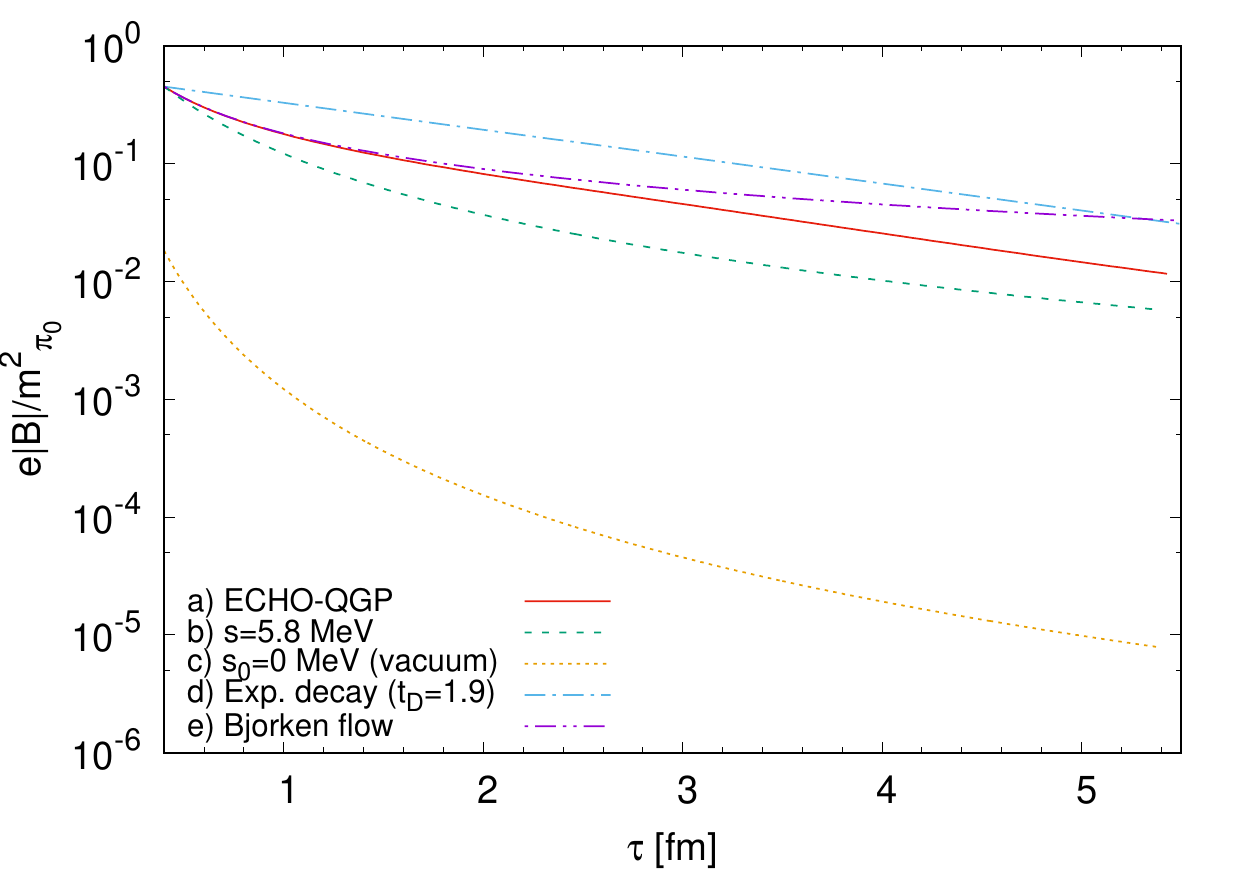}
\caption{
Time evolutions of magnitude of the magnetic fields at the central
 point of the collision. 
Taken from \cite{Inghirami:2016iru}. 
}
}
\label{fig:b-evolution}
\end{figure}

In most of the calculations so far, the sources of electromagnetic fields are
treated as classical ones.
The importance of quantum effects in
the estimation of magnetic fields is pointed out in
Ref.~\cite{Peroutka:2017esw}
The authors treated the sources as wave
packets satisfying the Dirac equation, and the obtained field
configurations turned out to show different behavior from classical
treatment.

\subsection{Vorticity}

Formation of vortices in heavy-ion collisions is attracting renewed
attentions since the report of finite $\Lambda$ polarization from STAR
\cite{STAR:2017ckg}.
The fluids formed after collisions naturally have
global vortical structure pointing in the direction of the angular
momentum.
In addition, in event-by-event initial conditions of fluid calculated
from transport models (HIJING/UrQMD), more complex vortex structures are
found \cite{Jiang:2016woz,Deng:2016gyh,Pang:2016igs,Karpenko:2016jyx}. 
Since the life-time and magnitude of vorticities are less uncertain
compared to magnetic fields, detection of CVE in heavy-ion collisions
can be more feasible than CME.
Since vorticities are larger at lower collisions energies, the
search for CVE can benefit from the Beam Energy Scan II at RHIC. 
Event-by-event anomalous hydrodynamic analysis of $\gamma$
correlation has been reported \cite{xingyu2016} in this conference.

\subsection{Isobaric collisions}

$\gamma$ correlation can be contaminated with background effects, 
such as transverse momentum conservation
\cite{Bzdak:2010fd}, charge conservation \cite{Schlichting:2010na}
and cluster particle correlations \cite{Wang:2009kd}. 
Those effects are ``flow driven'' because their contributions
are proportional to $v_2$. 
In order to identify the contributions from anomalous transport, 
RHIC is planning to perform the collisions of isobars using 
$ ^{96}_{44}{\rm Ru}$ and $ ^{96}_{40}{\rm Zr}$ in 2018.
Since those isobars have the same mass number, the geometry of the
collisions of ${\rm Ru}+{\rm Ru}$ and ${\rm Zr}+{\rm Zr}$ are the same. 
But the numbers of protons are different and the strength of
the magnetic fields can be varied without changing the flow.
In Ref.~\cite{Deng:2016knn}, it is shown that the two types of collisions can
indeed give rise to sizable ($\sim 20 \% $) difference in the
observables.
Figure 5
shows
$\langle B^2 \cos \left[2 (\Psi_{\rm B} -
\Psi_{\rm RP})\right] \rangle $
as a function of centrality for the two
types of collisions.
The quantity 
$\langle B^2 \cos \left[2 (\Psi_{\rm B} -
\Psi_{\rm RP})\right] \rangle $
is a good measure for the anomalous contribution for the following
reasons: 
$\gamma$ correlation from anomalous transport should scale as $|\bm B|^2$. 
Since $\gamma$  quantifies charge separation in the out-of-plane
direction, if the direction of $\bm B$ field ($\Psi_{\rm B}$) is
decorrelated with $\Psi_{\rm RP}$, the signal should vanish.
The quantity $\langle B^2 \cos \left[2 (\Psi_{\rm B} -
\Psi_{\rm RP})\right] \rangle $ captures this. 
\begin{figure}[htbp]
 \center{
\includegraphics[width=0.6\textwidth]{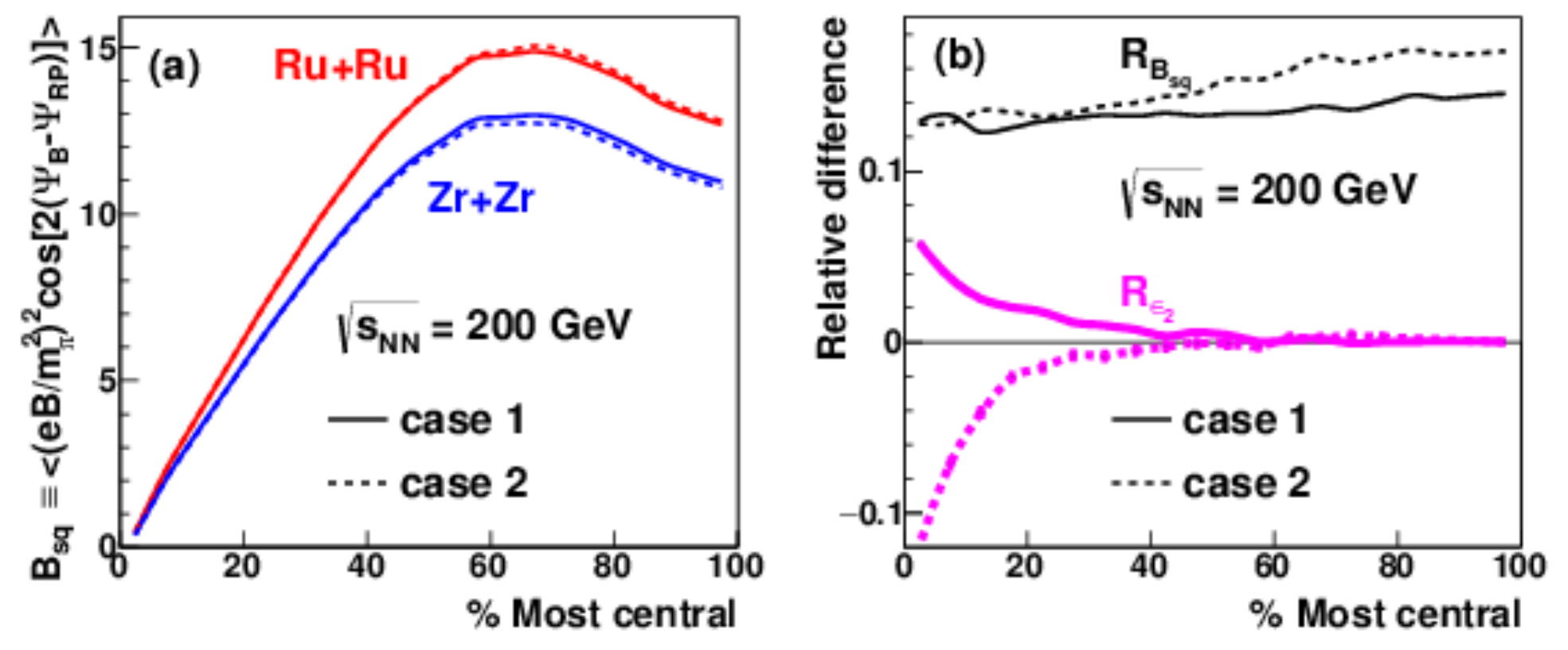}
 \caption{
 Comparison of
 $\langle B^2 \cos \left[2 (\Psi_{\rm B} -
 \Psi_{\rm RP})\right] \rangle $
 between ${\rm Ru}+{\rm Ru}$ and  ${\rm Zr}+{\rm Zr}$  collisions
Taken from Ref.~\cite{Deng:2016knn}. 
 }
}
\label{fig:isobar}
\end{figure}

\section{Summary}

In summary, the interplay of chiral fluids and dynamical
electromagnetic fields leads to a rich variety of phenomena. 
The fermions affect the topological configuration of magnetic fields and
fluid velocities. 
There are ongoing efforts for more sophisticated description of
anomalous chiral effects aiming at the detection of those effects in
heavy-ion collisions.

\section*{Acknowledgements}

This work  supported by the 
U.S. Department of Energy, Office of Science, Office of Nuclear Physics,
under contract No. DE-SC0012704 and within the framework of the Beam
Energy Scan Theory (BEST) Topical Collaboration.

%
%
%
%
%
%
%
%
%





\bibliographystyle{elsarticle-num}
\bibliography{refs.bib}







\end{document}